          [2001/04/25 1.1 (PWD)]
\documentclass[a4paper,twocolumn]{esapub2005} % European paper
\usepackage{times}
\usepackage{natbib}
\usepackage{graphicx}

\def\etal{{\it et~al.}}
\def\a{$^{\mbox{\scriptsize a}}$}
\def\b{$^{\mbox{\scriptsize b}}$}
\def\c{$^{\mbox{\scriptsize c}}$}
\def\d{$^{\mbox{\scriptsize d}}$}

\title{Spectral analysis of X-ray pulsars with the INTEGRAL observatory}
\author{E.V. Filippova}
\author{S.S. Tsygankov}
\author{A.A. Lutovinov} 
\author{R.A. Sunyaev}
\affil{Space Research Institute, Profsoyuznaya str. 84/32, Moscow 117997, Russia}

\begin{document}


\maketitle

\begin{abstract}
We studied spectra for 34 accretion-powered X-ray and one millisecond pulsars that were within
the field of view of the INTEGRAL observatory over two years (December 2002--January 2005) of its
in-orbit operation and that were detected by its instruments at a statistically significant level ($>$ 8$\sigma$ in
the energy range 18--60 keV). There are seven recently discovered objects of this class among the pulsars
studied: \hbox{2RXP J130159.6-635806}, \hbox{IGR/AX J16320-4751}, \hbox{IGR J16358-4726}, AX J163904-4642,
\hbox{IGR J16465-4507}, SAX/IGR J18027-2017 and AX~J1841.0-0535.
We analyze the evolution of spectral
parameters as a function of the intensity of the sources and compare these with the results of previous
studies.
\end{abstract}

\section{Introduction}

There are about 100 accretion-powered 
X-ray pulsars  are known to date and many paper have been devoted as to the 
study of individual sources as to the reviews ones. Reference \cite{w1983}  
were the first, who summarized the
spectra and pulse profiles for the 
X-ray pulsars and proposed an empirical model to
describe their spectra. Subsequently, reference \cite{n1989}
gave an overview of accretion-powered pulsars.  
GRANAT (\cite{l1994}) and ComptonGRO (\cite{b1997})  data  were used to investigate the
pulse profiles and the evolution of the
pulse periods. Reference \cite{c2002}
 used the RXTE data to analyze the pulsars whose spectra exhibited
the cyclotron lines.
  
Here we shortly report results of the broad band spectral analysis of the 
X-ray pulsars  observed by the INTEGRAL observatory (\cite{w2003}).
The full review with detailed description of results for individual 
sources with corresponding references can be found in \cite{f2005}.

\section{Observations}
In this work  we used the INTEGRAL observations from
orbit 23 (MJD 52629, December 21, 2002) to
orbit 239 (MJD 53276, September 28, 2004); these
were the publicly available at that moment data and the
data of the Russian quota obtained as part of the
Galactic plane scanning (GPS), the Galactic center
deep exploration (GCDE), and the observations in
the General Program. The publicly available
observations of the X-ray pulsar V0332+53 that were
performed from orbit 272 (MJD 53376, January 6,
2005) to orbit 278 (MJD 53394, January 24, 2005)
were used as an exception. Data from the
ISGRI detector of the IBIS telescope and from the JEM-X monitor were 
used for the analysis.

\section{Data Analysis}

     For all of the detected X-ray pulsars  we
constructed light curves in the energy range $18-60$ keV
and analyzed their variability. We constructed
average spectra for persistent sources and analyzed the
dependence of the spectrum on the source's state
for pulsars with variable fluxes:
 if the spectrum did
not change, we also provided an average spectrum;
otherwise, we gave the spectra of different states.
To fit the spectra, we used a standard (for pulsars)
empirical model that includes a power law with a
high-energy cutoff (\cite{w1983}).
 In certain cases, the standard model did not
describe the pulsar's spectral shape quite accurately.
Therefore, we introduced some additional components
when fitting the spectrum:
 low-energy photoelectron absorption,
 an iron emission line described by a Gaussian
profile,
 a resonance cyclotron scattering feature.

\begin{table*}

\caption{List of detected with INTEGRAL X-ray pulsars and their best-fit spectral parameters}
\begin{small}
\begin{tabular}{l|c|c|c|c|c}
\hline
\hline
&&&&&\\
Name&$N_{H}$,10$^{22}$ sm$^{-2}$&Photon index, $\Gamma$& $E_{cut}$,keV & $E_{fold}$, keV & $\chi^{2}$ \\
&&&&&\\
\hline
A 0114+650  &  --& 2.3$\pm$0.4& --& --& 0.42(6)\\[1mm]
SMC X-1 &-- &1.48$\pm$0.02& 20.5$^{+1.0}_{-1.8}$&12.9$^{+0.6}_{-0.7}$&0.98(124) \\[1mm]
RX J0146.9+6121&--&$2.9^{+1.1}_{-0.8}$&--&--&0.31(3)\\[1mm]
V0332+53 &4\a&$0.77\pm0.02$&$24.3^{+0.5}_{-0.7}$&$14.0^{+0.5}_{-0.7}$&0.35(127)\\[1mm]
4U 0352+309 & -- &1.92$\pm0.19$&50$\pm16$&77$\pm27$& 0.36(9) \\[1mm]
LMC X-4 &--&0.2$\pm0.15$& 9.1$\pm0.8$& 11.0$\pm0.6$& 0.93(117) \\[1mm]
A 0535+26& --& 1.2\a& 24\a&13.8$^{+4.5}_{-3.2}$&0.07(5)\\[1mm]
Vela X-1(eclipse) &--&3.1$\pm0.3$ &-- &--&0.83(7) \\[1mm]
Vela X-1(outside eclipse) &--&0.88$\pm0.01$ &25.5$\pm0.2$ &13.0$\pm0.1$&0.34(131) \\[1mm]
CEN X-3(quiescent state)& -- & 0.87$\pm0.06$ &16.4$\pm0.6$& 7.1$\pm0.2$ &1.5(120) \\[1mm]
CEN X-3(outbursts)&-- & 1.16$\pm0.04$ &15.3$\pm0.2$ & 7.8$\pm0.2$ &1.4(116) \\[1mm]
4U 1145-619 & -- & 1.5$\pm0.1$ & 6.7$\pm1.4$ & 30$\pm4$ & 1(142) \\[1mm]
1E 1145.1-614 &3.3\a & 1.08$\pm$0.07 & 8$\pm$1 & 21.9$^{+1.8}_{-0.8}$ &0.98(139) \\[1mm]
GX 301-2 (high state)& -- &0.74$^{+0.32}_{-0.09}$&23.3$^{+0.3}_{-0.5}$& 8.3$\pm0.7$& 0.74(8) \\[1mm]
GX 301-2 (low state)& 10.6$\pm2.5$ &0.30$\pm0.06$&17.8$\pm0.2$&9.7$\pm0.7$&0.9(118) \\[1mm]
2RXP130159.6-635806&2.56\a&0.69\a&24.3$\pm$3.4&8.5$^{+0.2}_{-0.1}$&\c\\[1mm]
4U 1538-52 & 1.63\a & 1.37$\pm0.06$ &28.7$\pm0.8$ & 9.9$\pm0.7$ &0.94(119) \\[1mm]
4U 1626-67 &-- & 0.87\a & 23.9$^{+1.0}_{-1.4}$ &7$\pm$1 &1.25(5) \\[1mm]
IGR/AX J16320-4752\b&18\a & 0.7$\pm$0.2& -- &13$\pm$1&\d\\ [1mm]
IGR J16358-4726\b &40\a&0.7$\pm$0.5& --& 16$\pm$5&\d\\[1mm]
AX J163904-4642\b &58\a& 1.3$\pm$1.0& -- & 11$\pm$1&\d\\[1mm]
IGR J16465-4507\b &72\a&1.0$\pm$0.5 & -- & 30\a&\d \\[1mm]
OAO 1657-415 &15.2$^{+0.7}_{-1.4}$ & 1.57$\pm0.02$& 26.3$^{+0.7}_{-1.8}$ & 29.2$^{+1.2}_{-0.5}$ &0.73(119) \\[1mm]
EXO 1722-363 &--&3.5\a& -- & -- &2.7(5) \\[1mm]
GX 1+4 (low state)&--&2.24$^{+0.06}_{-0.12}$ &-- &-- &0.93(126) \\[1mm]
GX 1+4 (intermediate state)&--&1.54$^{+0.35}_{-0.22}$ &24.8$^{+5.8}_{-3.0}$ &47.0$^{+15.2}_{-10.7}$ &1.16(125) \\[1mm]
GX 1+4 (high state)&--&$0.93^{+0.12}_{-0.14}$&$25.1^{+1.1}_{-1.7}$&$30.4\pm2.4$&1.19(136) \\[1mm]
IGR/SAX J18027-2017\b &--&0.1\a&--&$\sim$10&-- \\[1mm]
XTE J1807-294& -- & 1.96\a& 48.1$^{+7.6}_{-9.9}$& 75.7$^{+58.1}_{-24.5}$& 0.92(7) \\[1mm]
AX J1820.5-1434 & -- & 0.9\a & 25$\pm3$ & 17.0$\pm2.7$ &0.37(9) \\[1mm]
AX J1841.0-0535&-- &$2.2\pm0.3$& --&--&0.42(5)\\[1mm]
GS 1843+009&--&0.34\a&5.95\a& 17.4$\pm1.4$&1.2(8) \\[1mm]
A 1845-024& -- & 2.62$\pm0.19$ & -- &--&0.46(7)\\[1mm]
XTE J1855-026 & --&1.69$\pm0.23$&23.99$^{+2.88}_{-6.73}$&38.49$^{+10.35}_{-7.38}$&1.08(112)\\[1mm]
XTE J1858+034 &14.3$\pm0.7$&1.38$\pm$0.02&25.16$\pm0.33$&7.92$\pm$0.22&0.95(144)\\[1mm]
X 1901+031&--& 2.035$\pm0.015$ & 11.27$\pm0.19$ & 13.22$\pm0.11$ & 0.82(127) \\[1mm]
4U 1907+097 &-- & 1.26$\pm0.07$ & 7.0$\pm0.3$ &9.0$^{+0.3}_{-0.6}$ &0.75(131) \\[1mm]
KS 1947+300&--& 1.07$^{+0.24}_{-0.13}$&8.6$^{+3.4}_{-1.2}$& 23.6$^{+5.3}_{-2.3}$& 1.18(104) \\[1mm]
EXO 2030+375 &-- & 1.71$\pm0.09$ &25.2$^{+2.5}_{-3.7}$ & 33$^{+6}_{-4}$ &1.06(137) \\[1mm]
SAX J2103.5+4545 & 0.9\a& 1.04$\pm0.15$ & 8.5$\pm2.4$ & 21.37$\pm2.75$ &1.21(120) \\[1mm]
\hline
\end{tabular}
\end{small}

\begin{scriptsize}
\a  The parameter is fixed\\
\b  The cutoffpl model was used to fit the spectrum\\
\c  The pulsar's spectrum was fitted over a
wide energy range
together with data from XMM observatory \cite{c2005} .\\
\d   The pulsar's spectrum was fitted over a
wide energy range
together with data from XMM observatory \cite{l2005}.\\

\end{scriptsize}

\end{table*}

\begin{figure*}
\hbox{
\includegraphics[width=0.7\columnwidth,bb=30 435 565 710,clip]{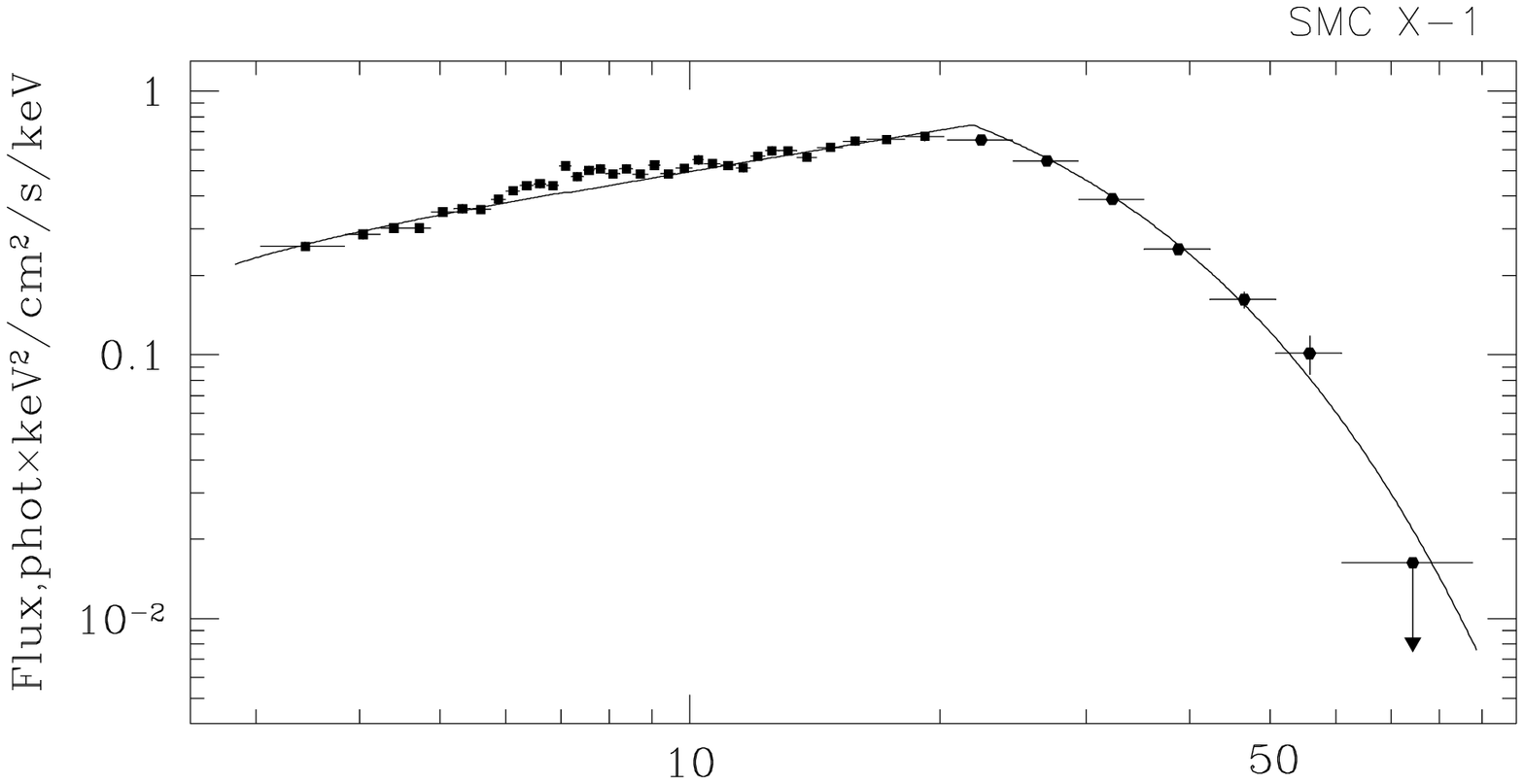}
\includegraphics[width=0.7\columnwidth,bb=30 435 565 710,clip]{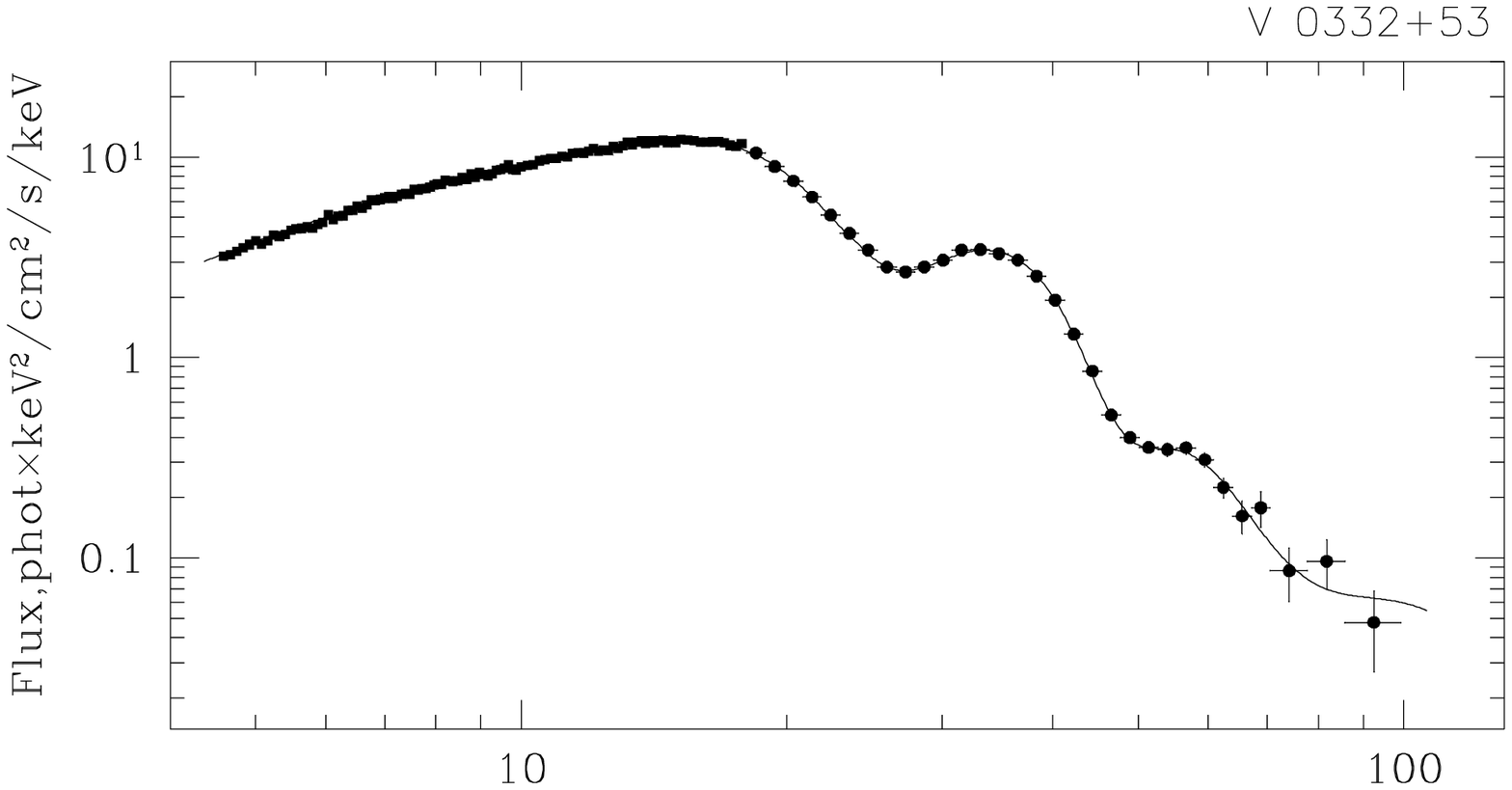}
\includegraphics[width=0.7\columnwidth,bb=30 435 565 710,clip]{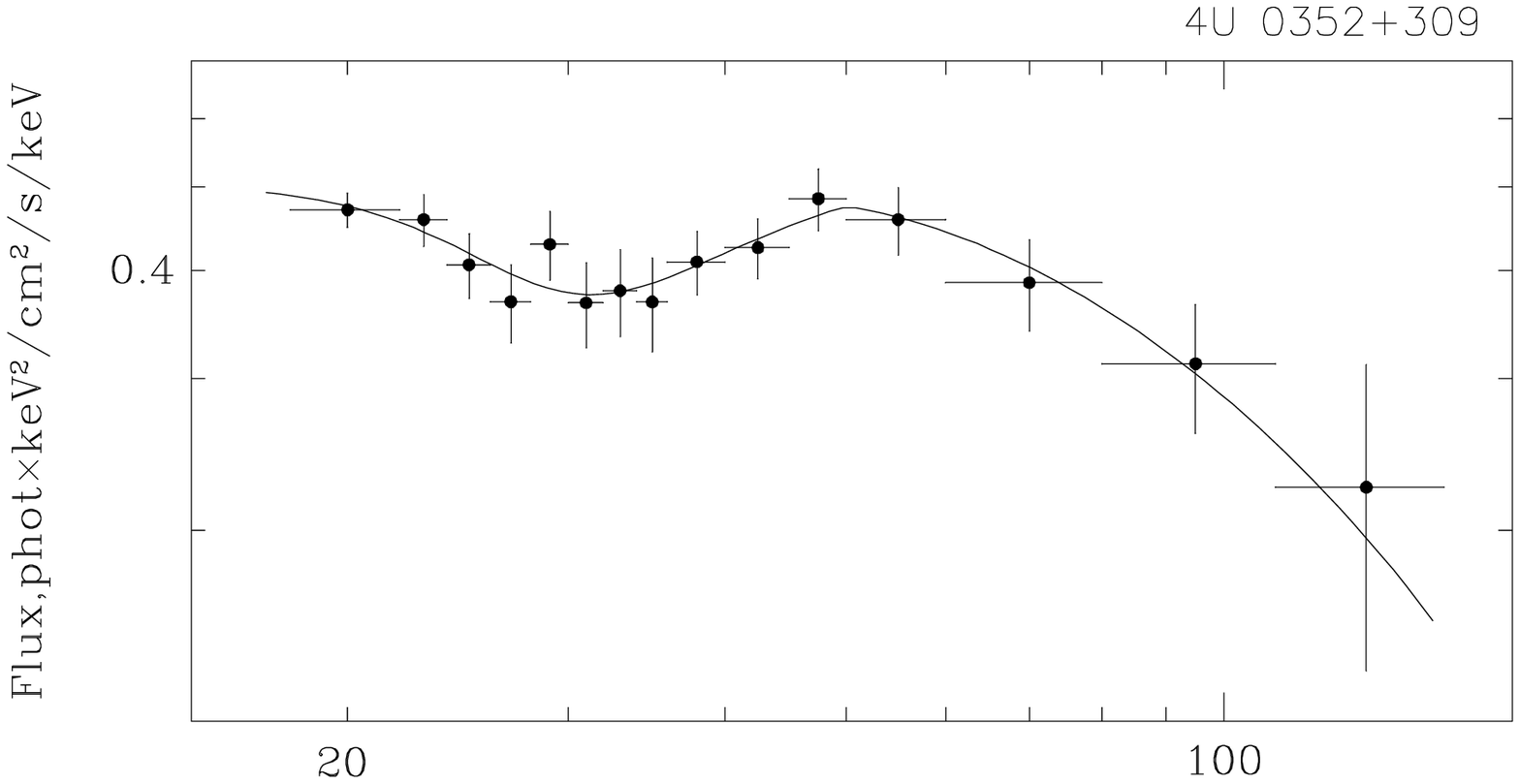}
}
\hbox{
\includegraphics[width=0.7\columnwidth,bb=30 435 565 710,clip]{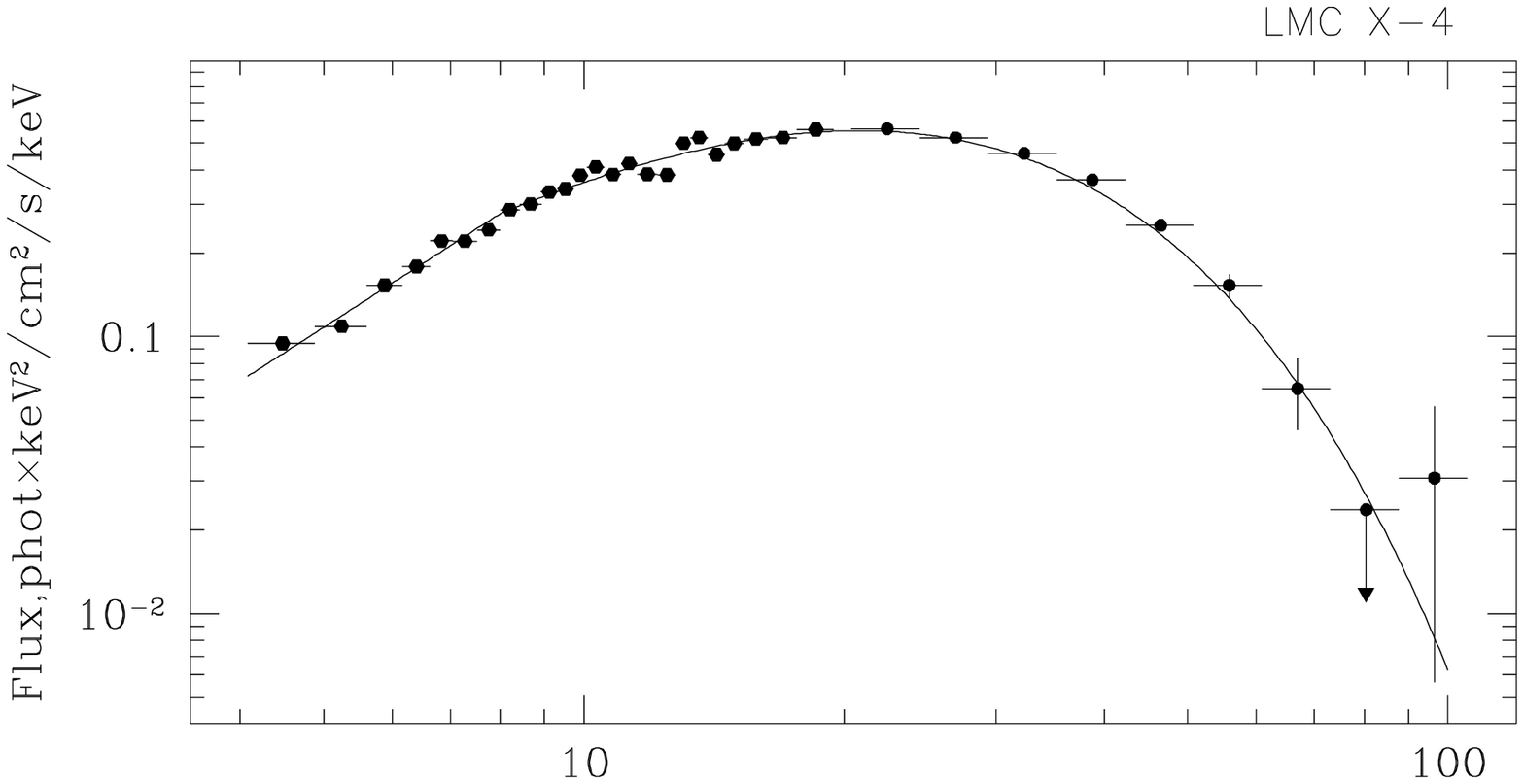}
\includegraphics[width=0.7\columnwidth,bb=30 410 565 710]{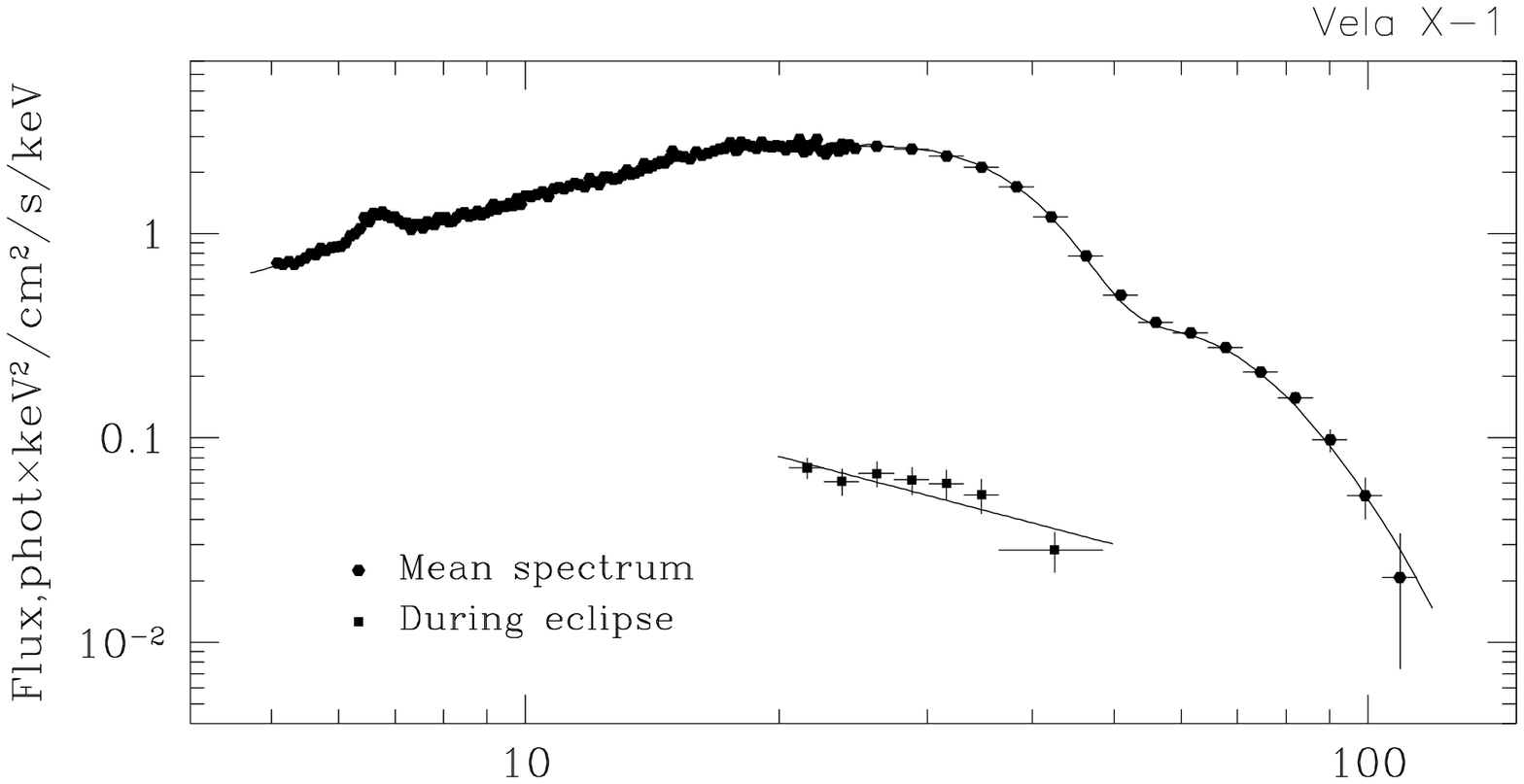}
\includegraphics[width=0.7\columnwidth,bb=30 410 565 710]{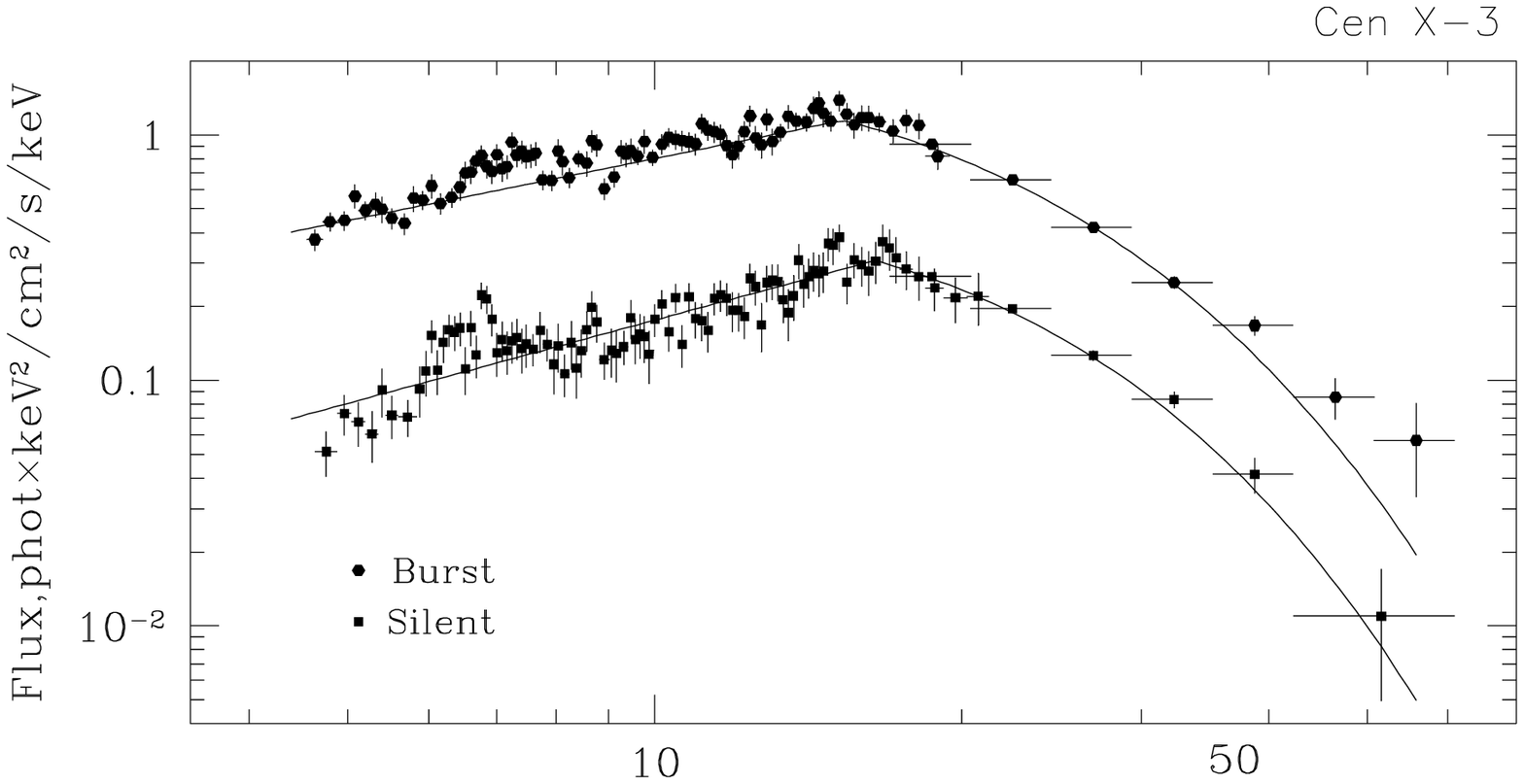}
}
\hbox{
\includegraphics[width=0.7\columnwidth,bb=30 410 565 710]{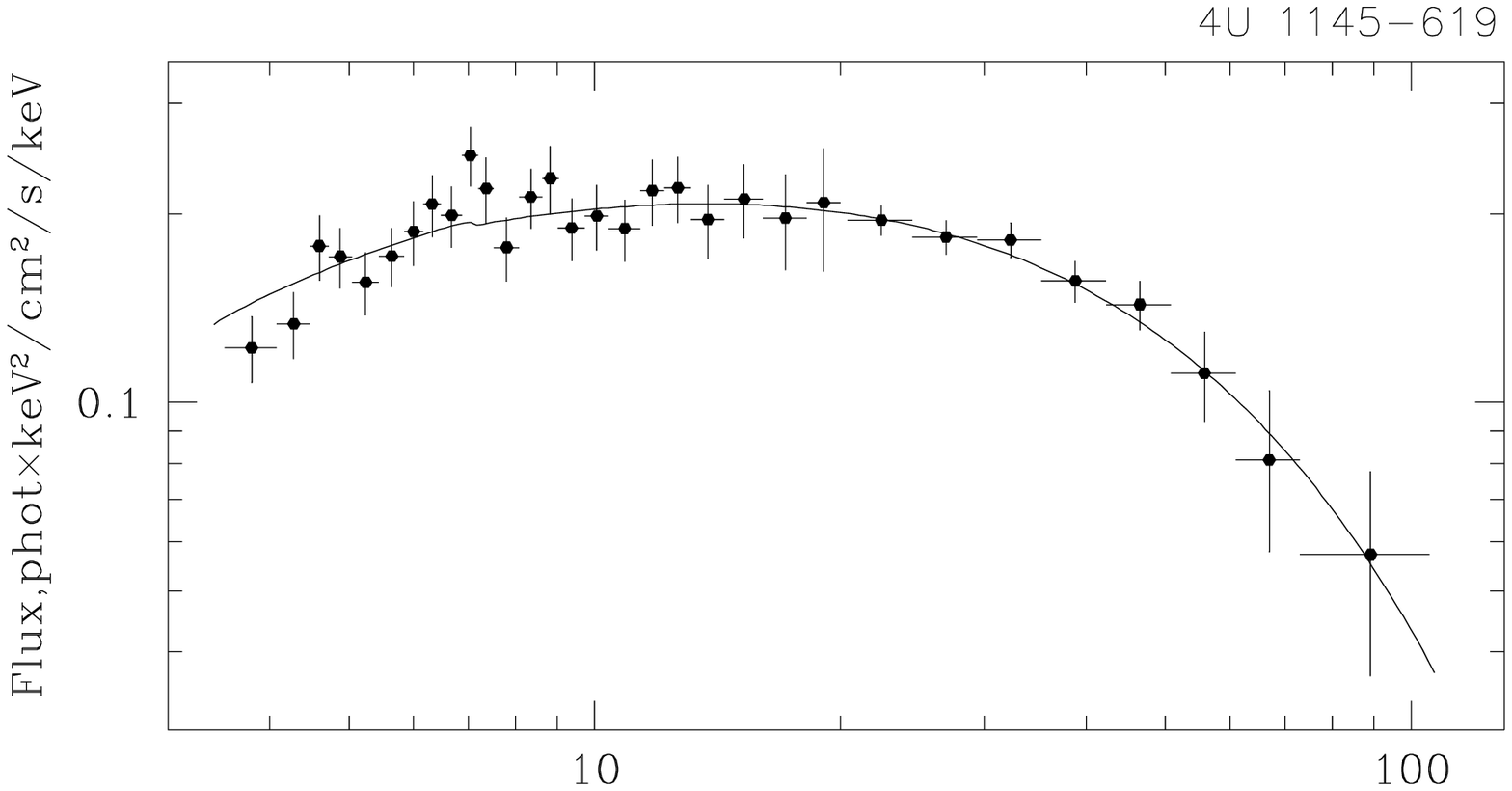}
\includegraphics[width=0.7\columnwidth,bb=30 410 565 710]{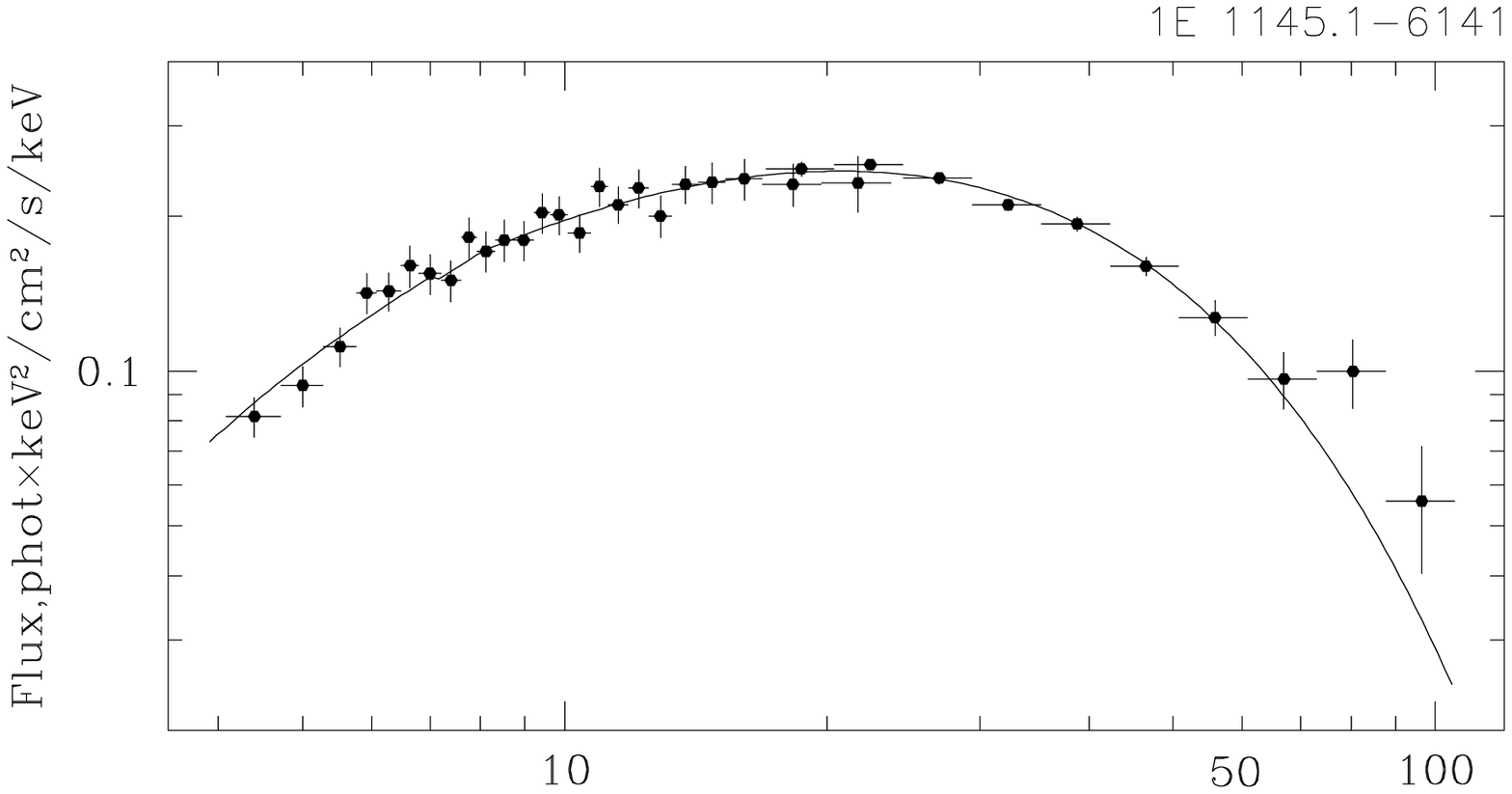}
\includegraphics[width=0.7\columnwidth,bb=30 410 565 710]{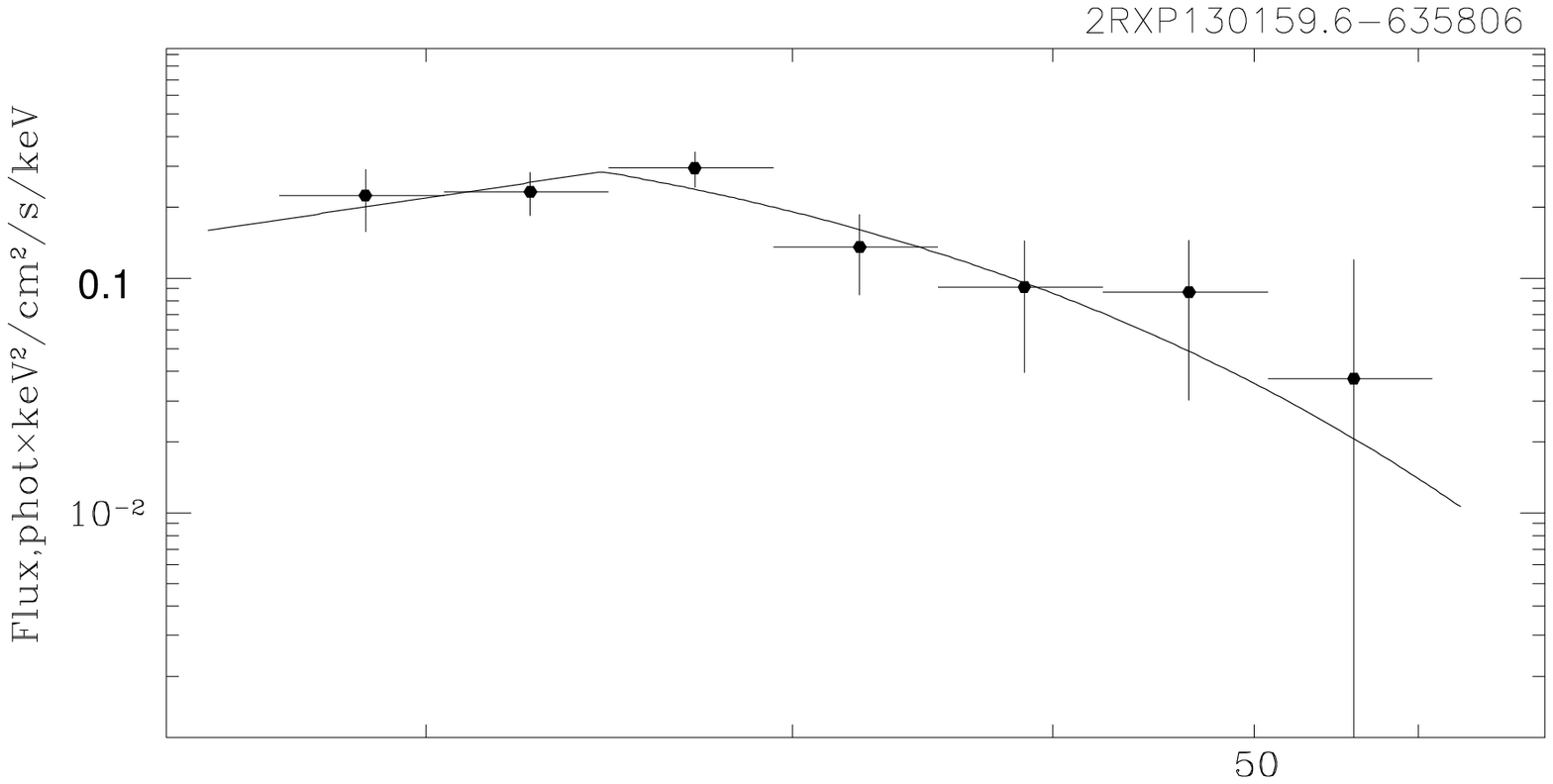}
}
\hbox{
\includegraphics[width=0.7\columnwidth,bb=30 410 565 710]{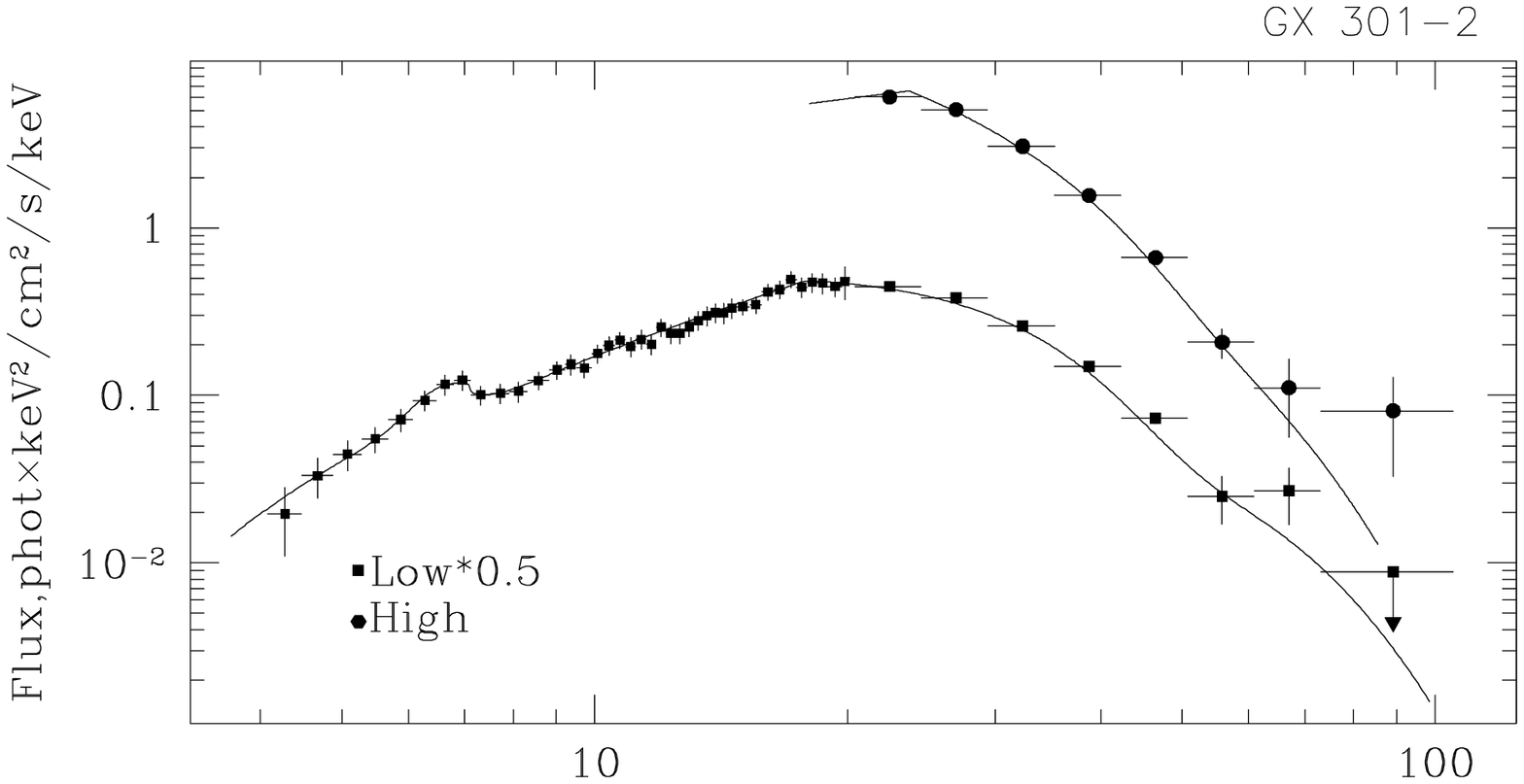}
\includegraphics[width=0.7\columnwidth,bb=30 410 565 710]{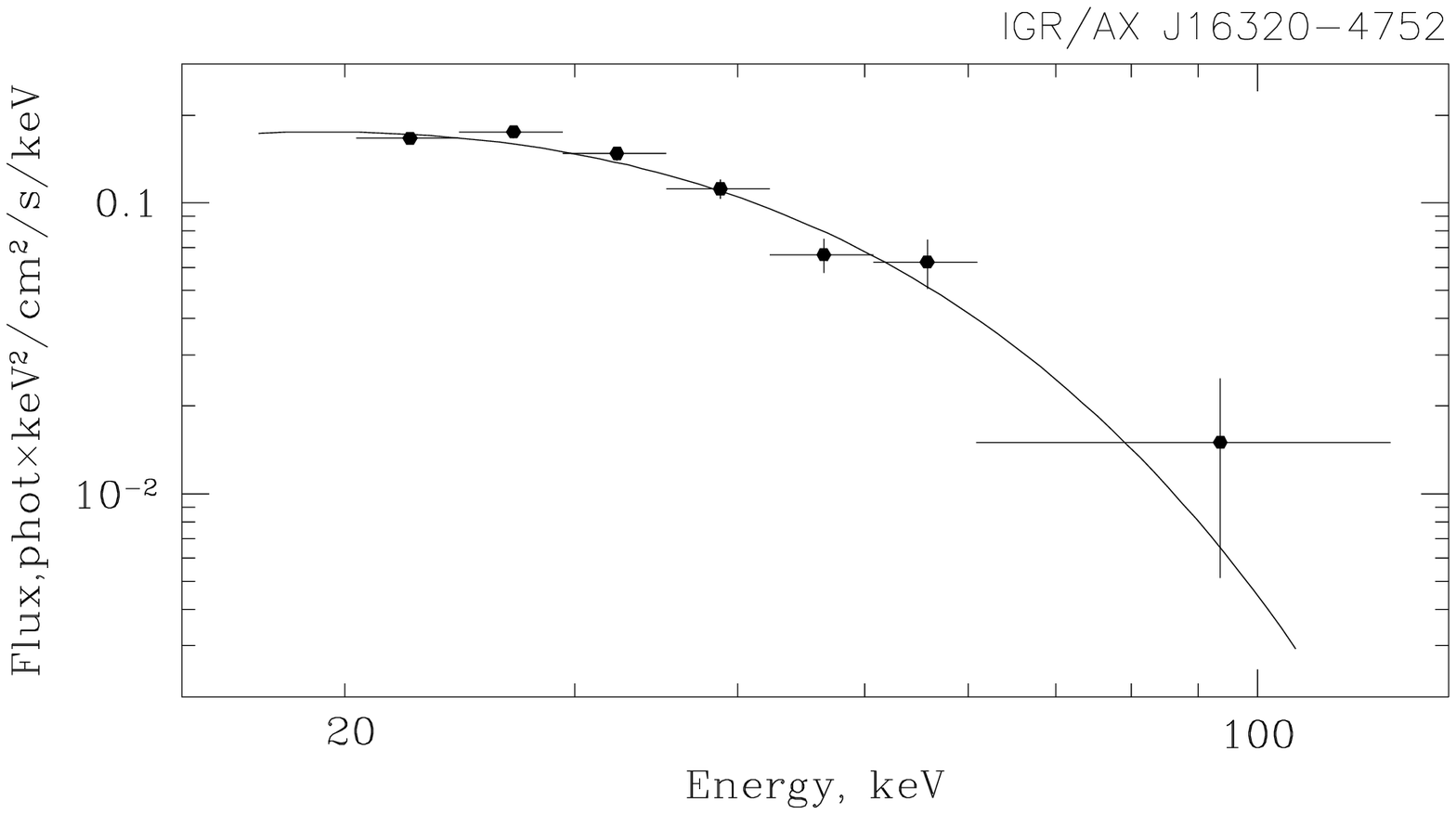}
\includegraphics[width=0.7\columnwidth,bb=30 410 565 710]{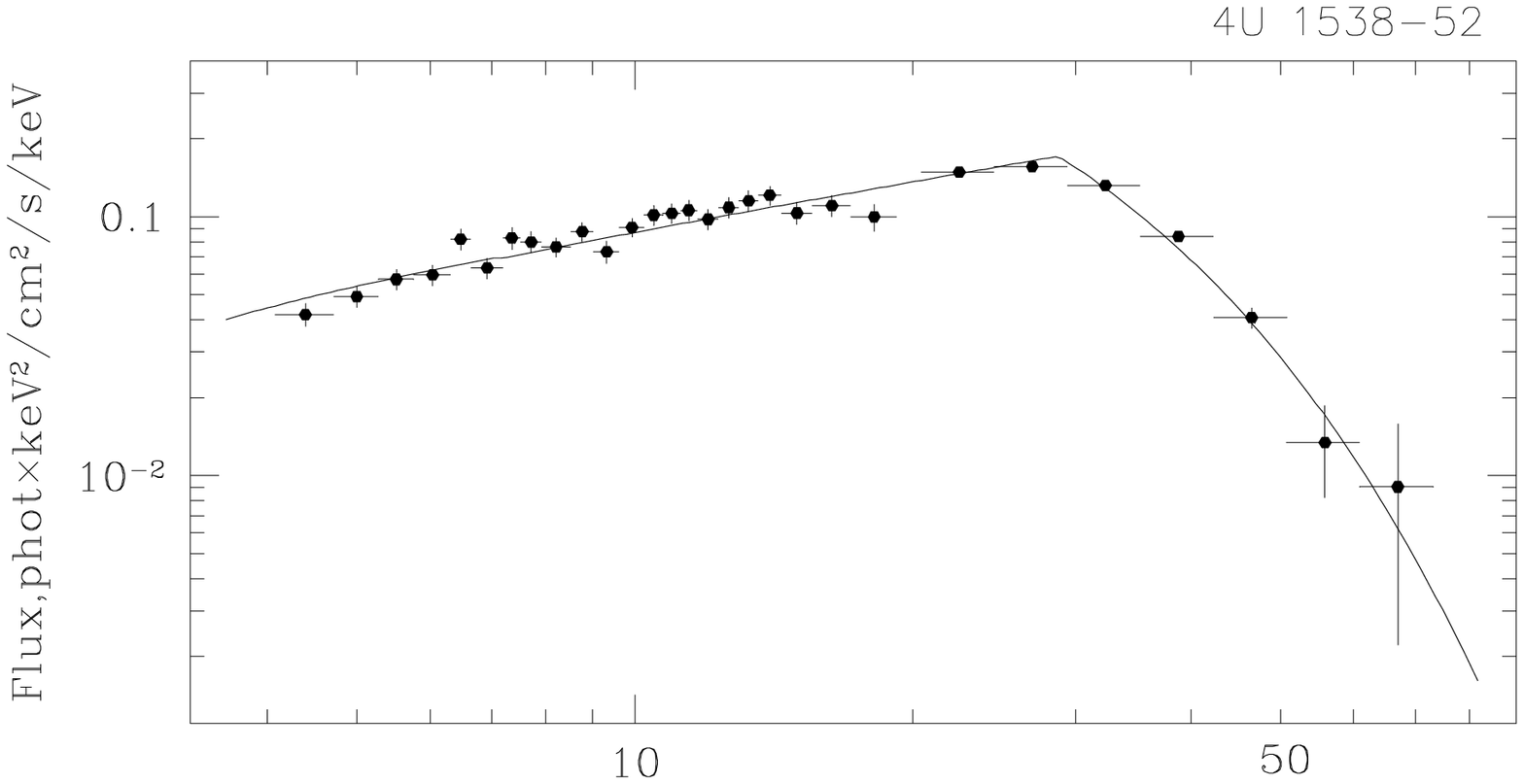}
}
\hbox{
\includegraphics[width=0.7\columnwidth,bb=30 410 565 710]{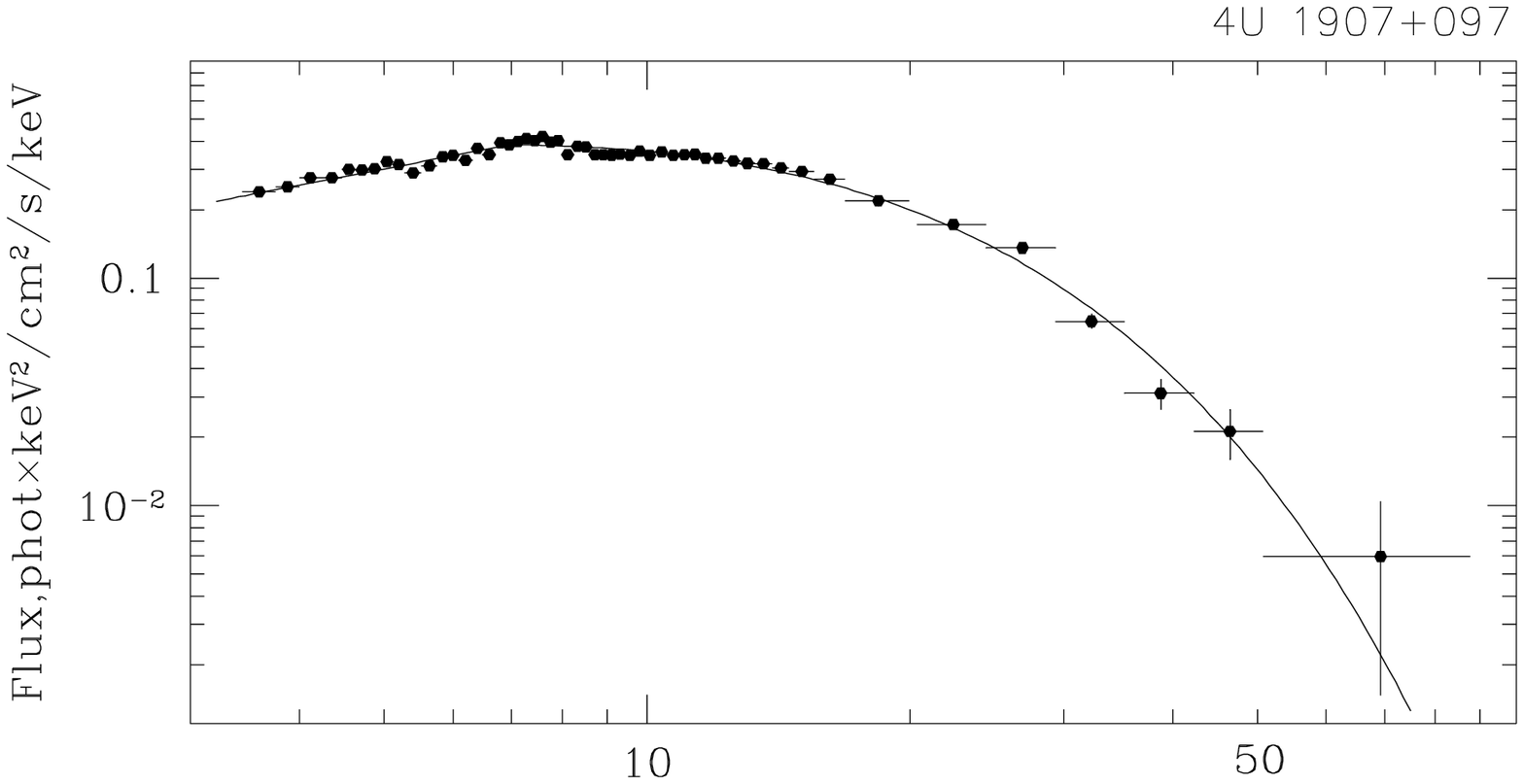}
\includegraphics[width=0.7\columnwidth,bb=30 410 565 710]{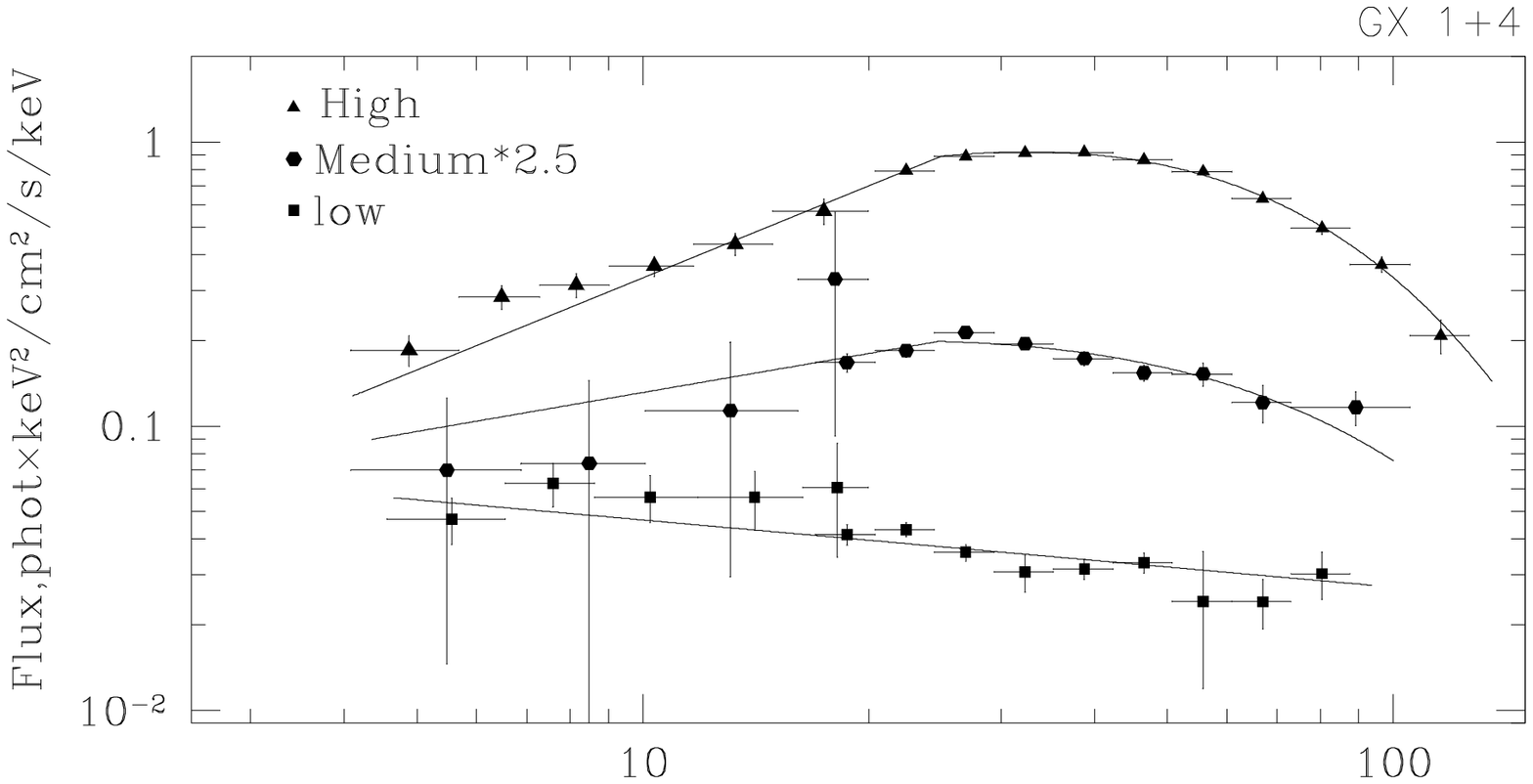}
\includegraphics[width=0.7\columnwidth,bb=30 410 565 710]{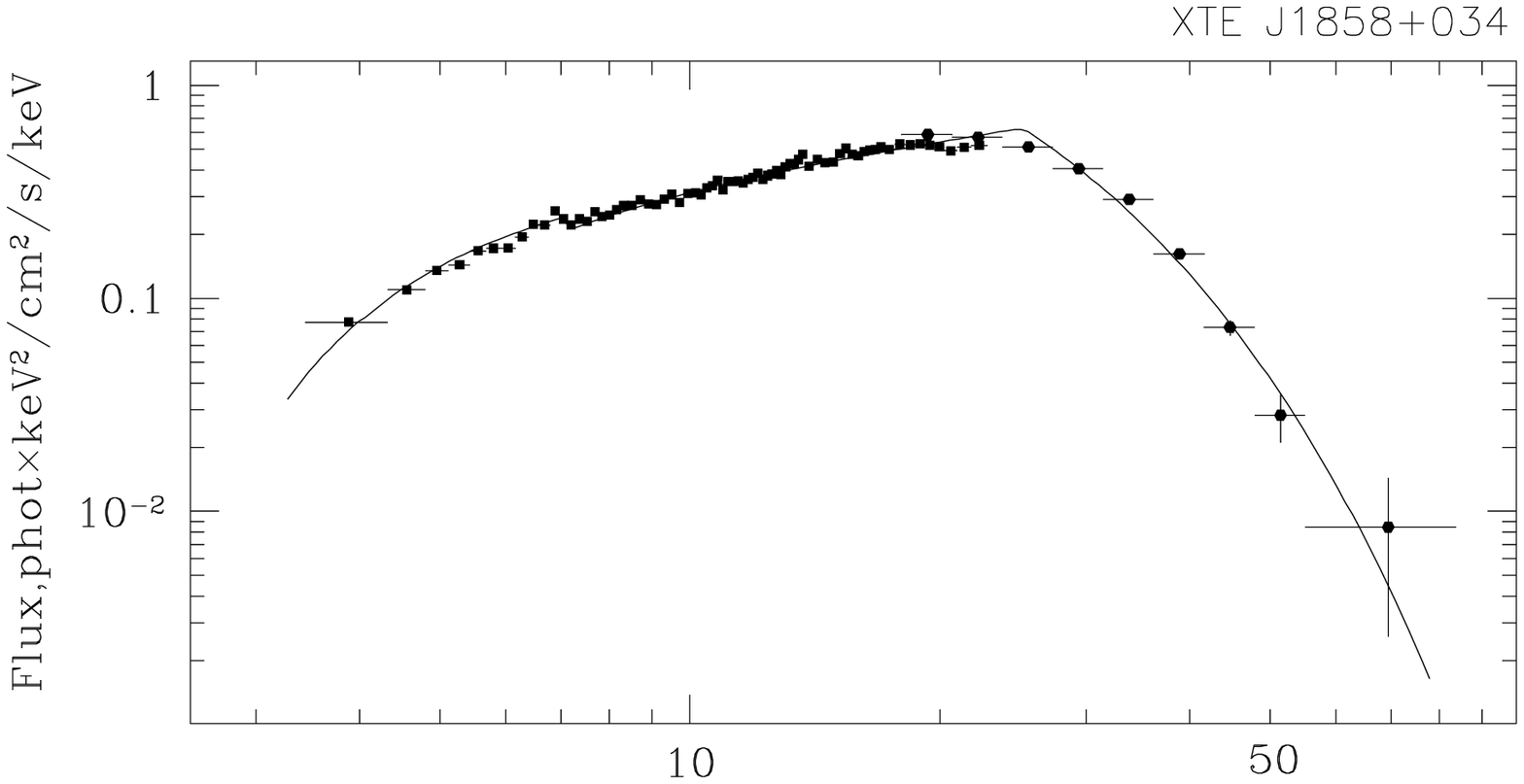}
}
\hbox{
\includegraphics[width=0.7\columnwidth,bb=30 410 565 710]{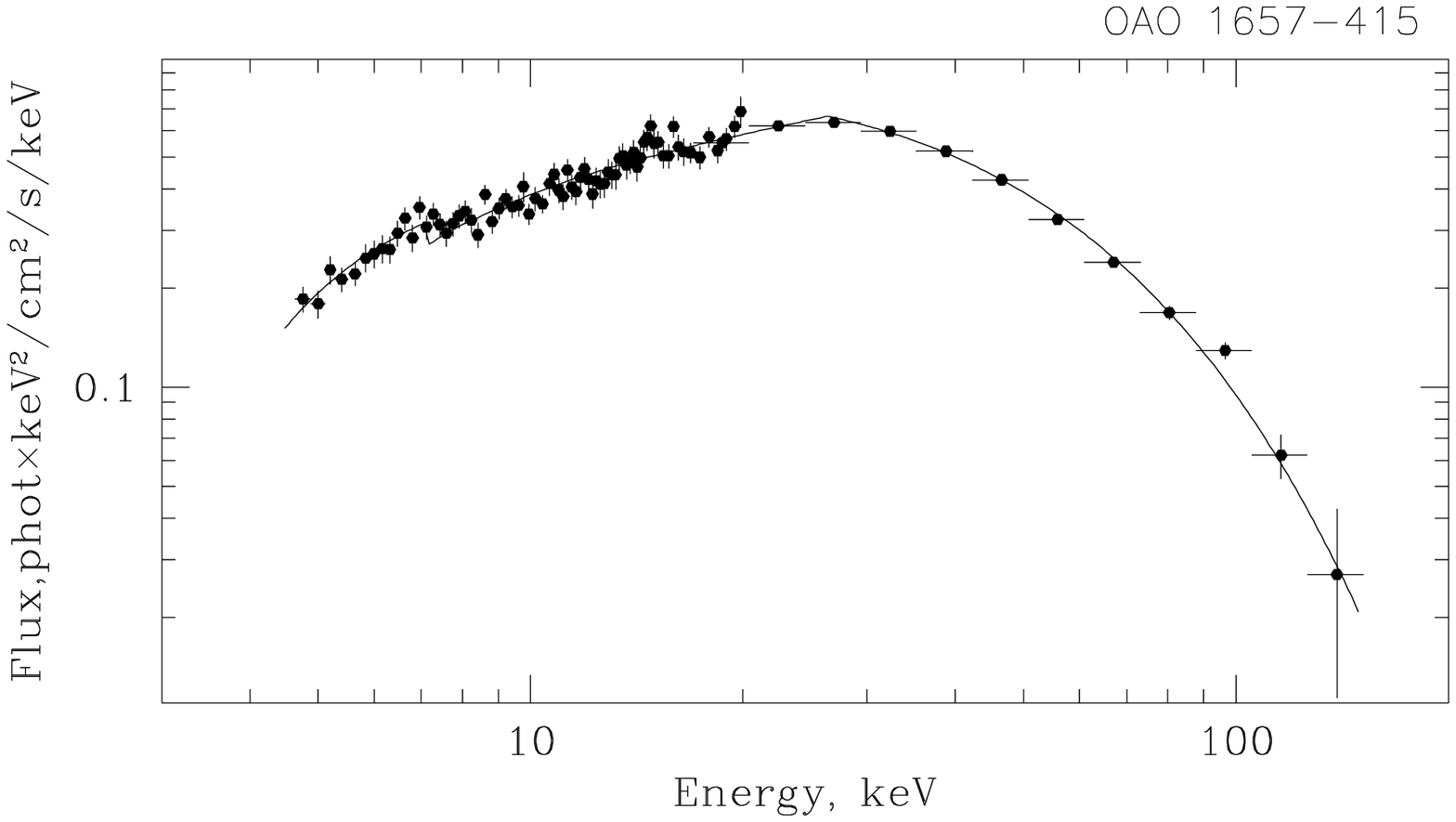}
\includegraphics[width=0.7\columnwidth,bb=30 410 565 710]{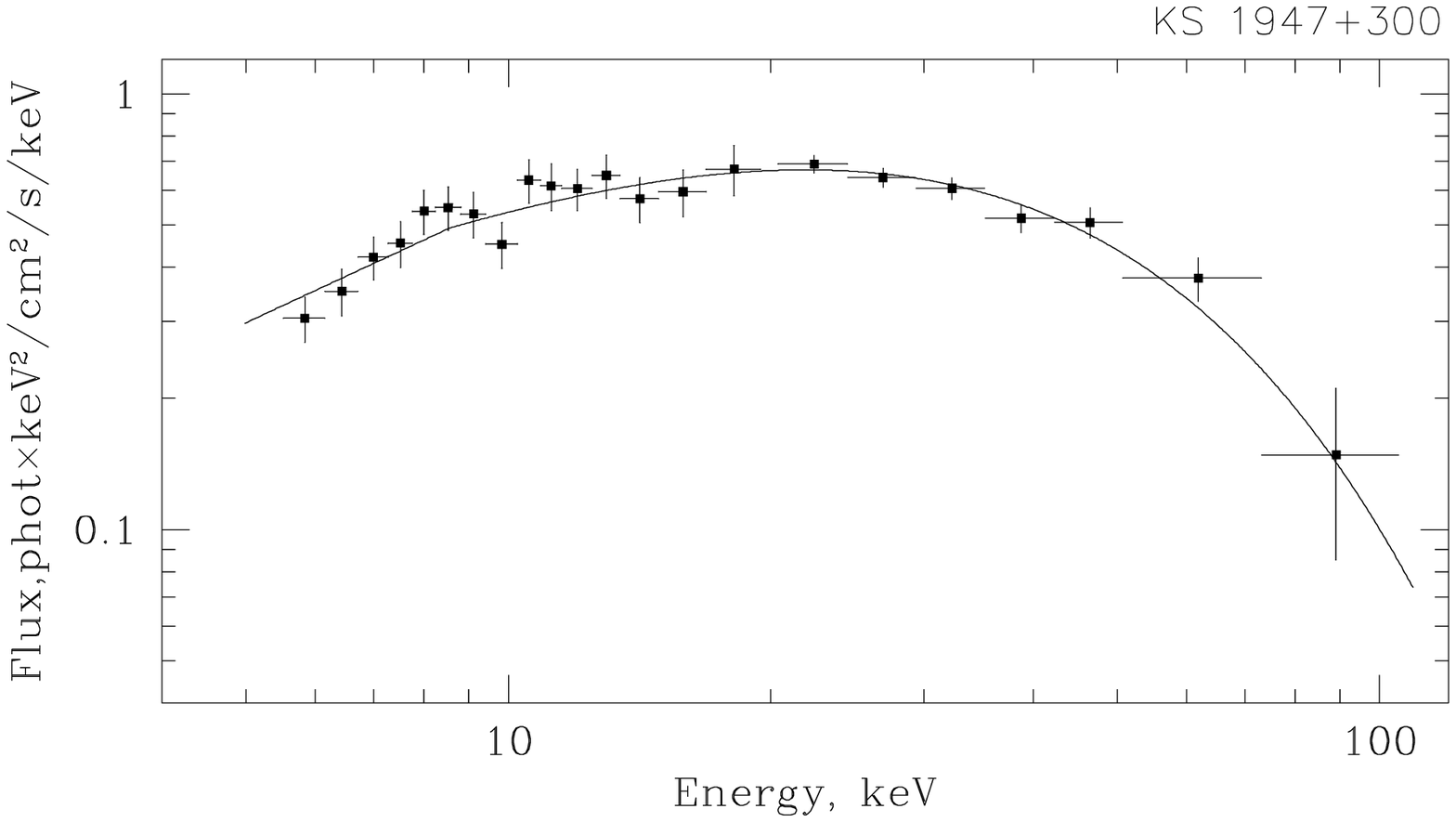}
\includegraphics[width=0.7\columnwidth,bb=30 410 565 710]{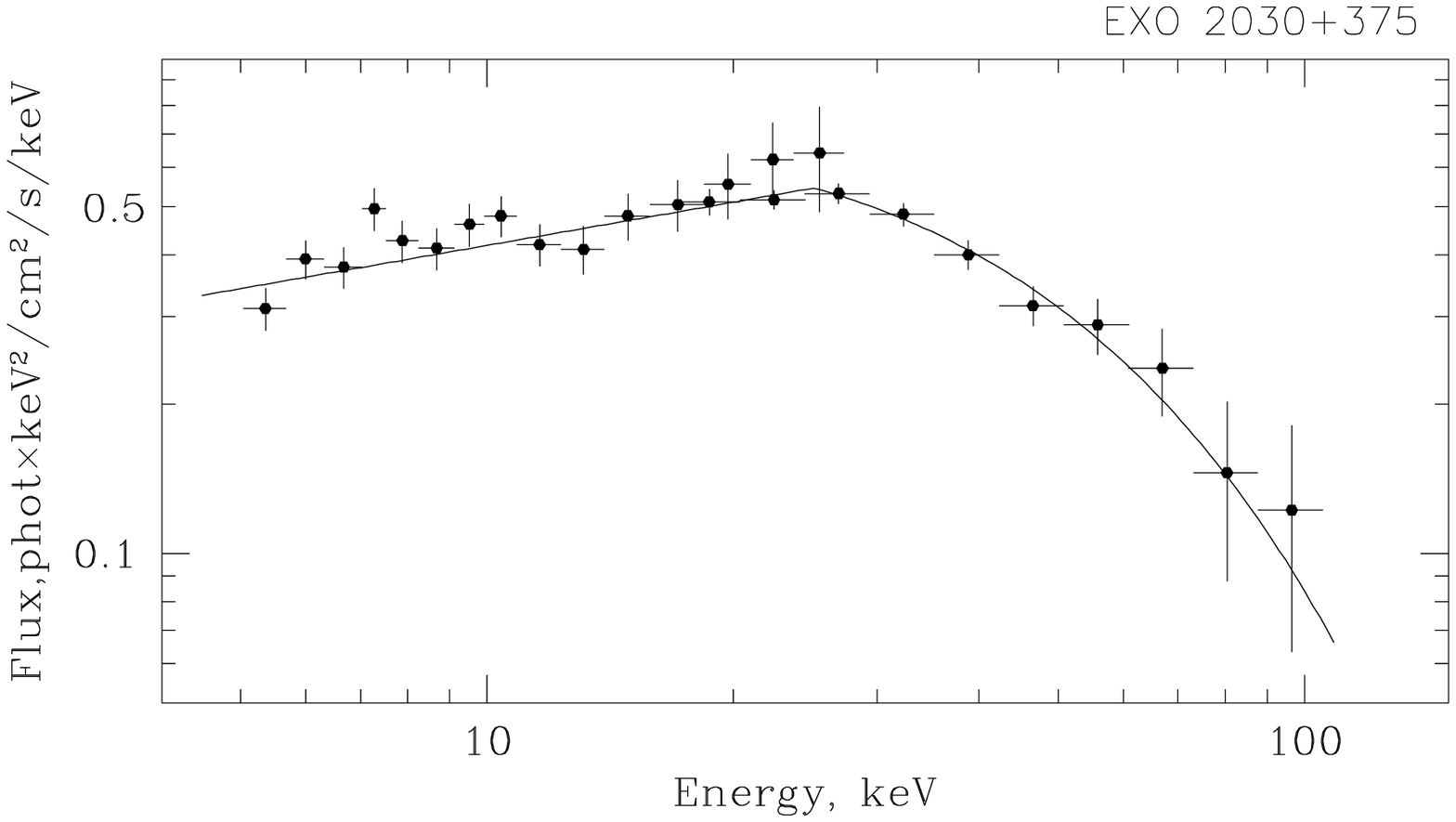}
}
\caption{INTEGRAL energy spectra for 18 X-ray pulsars. The solid lines
represent the best fit to the spectrum. The errors
correspond to one standard deviation.}
\end{figure*}

\section{Results}
As it was mentioned above we constructed and studied spectra of 35 sources. List of detected 
pulsars with best fit parameters  of their spectra are presented in Table 1. 
A couple of broadband spectra ($\sim 4-100$ keV) of 18 pulsars 
are shown in Fig.1. Below we briefly dwell on the mostly interesting results.

{\it First detection of the hard X-ray emission from several sources.}
Hard X-ray spectra for X-ray pulsars RX~J0146.9+6121, AX~J1820.5-1434 and AX~J1841.0-0535 
have been obtained for the first time.
The  pulsar RX~J0146.9+6121 is a faint source with the 18 -- 60 keV flux of 3 mCrab.
The pulsar AX~J1841.0-0535 was registered only during two outbursts, when its 18 -- 60 keV flux 
was increasing to  $\sim10-40$ mCrab.
Because of the low statistics data for these pulsars we
used a simple power law with the photon
index $\Gamma=2.9^{+1.1}_{-0.8}$  and  $\Gamma=2.2\pm0.3$, respectively, to fit their spectra.
For the pulsar AX~J1820.5-1434 the statistics was relatively good (the source was detected at a statisticaly significant level up to 
$\sim$ 70 keV), therefore we used the model 
of powerlaw with the high-energy cutoff to fit its spectrum.
The photon index was fixed at 0.9  taken from previous studies (\cite{k1998}).\\
We detected for the first time at a statistically significant 
level the 18 -- 60 keV flux of $\sim 7$ mCrab from the pulsar Vela X-1  during its eclipse by the optical companion. Since the
source was not detected by the JEM-X instrument
during the eclipse, we were able to construct its
spectrum only in the hard X-ray energy
range. The spectrum was fitted by a simple power law with the index
of 3.1$\pm$0.3 (Fig.1).

{\it Spectral variability of X-ray pulsars.}
The pulsar Cen X-3 is the eclipsing system; it demonstrates outbursts with the flux of $\sim90$ mCrab,
that is about 5 times higher than the source flux in the quiet state, $\sim$17 mCrab (in the 18 -- 60 keV energy range). 
We constructed the pulsar's radiation spectrum
averaged over all outbursts and an average persistent
spectrum outside the eclipses and found that the spectrum becomes softer during
outbursts: the photon index increases from 0.87 to 1.16.\\
 Our analysis confirmed (\cite{p1995}) that the spectral
parameters of the pulsar GX 1+4 radiation strongly depend on its flux.
Despite significant statistical errors (Fig.1) it is clear that as the intensity of the radiation from the
object under study decreases, its spectrum becomes
slightly softer. \\
Also we detected a statistically significant increasing in the photon index 
of the pulsar GX 301-2 during its transition from the low to the high state (Table 1). 

{\it Absorption and emission features.}
For several pulsars line features of a different nature were observed in their spectra.
We found three garmonics of the cyclotron absorption line in the spectrum of pulsar V 0332+53 at energies  
$E_{cycl1}$=24.25$^{+0.07}_{-0.14}$ keV, 
%$\tau_{cycl1}$=1.98$^{+0.02}_{-0.04}$, $\sigma_{cycl1}$=7.10$\pm$0.10,
E$_{cycl2}$=$46.8^{+0.2}_{-0.1}$ keV, 
%$\tau_{cycl2}$=1.94$^{+0.06}_{-0.07}$, $\sigma_{cycl2}$=8.9$\pm0.4$
E$_{cycl3}$=$67.9^{+3.2}_{-4.3}$ keV;
%, $\tau_{cycl3}$=2.60$^{+0.25}_{-0.35}$, $\sigma_{cycl3}$=26.9$\pm$5.4
two garmonics in the spectrum of pulsars Vela X-1 at energies 24.0$\pm0.3$ keV and $50.2\pm0.5$ keV;
one garmonic in the spectra of pulsars 4U 0352+309 and GX 301-2 at energies 28.8$\pm$2.5 keV and $\sim49$ keV, respectively.
It was found that for the pulsar V~0332+53 the 
cyclotron line energy is not constant but significantly changes with the luminosity (\cite{t2006})\\
A Fe line emission was detected in the spectra of the Vela X-1 and GX 301-2 (low state) at  energies
6.64$\pm0.10$ keV and  6.54$^{+0.17}_{-0.11}$ keV, respectively.
It worth mention that there are a number of
features near energies $5-7$ keV in spectra
reconstructed from the JEM-X data that are attributable to
the flaws in the current response matrix of the
instrument, that makes it difficult to identify the iron
emission line and to determine its parameters.

\section{Summary}

-- We constructed a catalog of spectra for 34
accretion-powered and one millisecond X-ray pulsars. Some of them were detected 
in hard X-rays for the first time. For 18 of the 35
sources, we were able to reconstruct their broadband
spectra. For variable sources, we analyzed the flux
dependence of the spectral shape.\\
-- A hard X-ray spectrum was obtained for the first time for the
pulsar \hbox{Vela X-1}  during an eclipse of
the source by its optical companion.\\
-- Cyclotron lines and their harmonics were detected
in the spectra of several pulsars: one harmonic in
4U~0352+309, one harmonic in both low and high
states in \hbox{GX~301-2}, two harmonics in Vela X-1, and
three harmonics in \hbox{V~0332+53}.

\section{Aknowledgments}
This work was supported by the Russian Foundation for
Basic Research (project no. 04-02-17276), the Russian Academy of Sciences
(The Origins and evolution of stars and galaxies program) and
grant of President of RF (NSh-1100.2006.2). AL acknowledges the financial
support from the Russian Science Support Foundation.

\end{document}